\documentclass[a4paper, oneside, twocolumn, notitlepage, 10pt]{extarticle_ecoc}
\usepackage{ecoc}

\begin{document}
\selectlanguage{english}    


\title{Simulation and Experimental Studies of DWDM Nonlinear Phase/Polarization/Power Crosstalk Between DFOS and Communication Channels in 27.6-Tb/s 800ZR Metro Network}


\author{
    Jingchuan Wang\textsuperscript{(1)}, Maoqi Liu\textsuperscript{(1,*)}, Liwang Lu\textsuperscript{(1)}, Alan Pak Tao Lau\textsuperscript{(1)}, Chao Lu\textsuperscript{(1)}
}

\maketitle                  
\vspace{-0.1cm}
\begin{strip}
    
    \begin{author_descr}
        \textsuperscript{(1)}Photonics Research Institute, The Hong Kong Polytechnic University, Hong Kong SAR, China,
        \textcolor{blue}{\uline{jingchuan98.wang@connect.polyu.hk, maoqi.liu@connect.polyu.hk}}
        
    \end{author_descr}
\end{strip}

\renewcommand\footnotemark{}
\renewcommand\footnoterule{}


\begin{strip}
    \begin{ecoc_abstract}
        We comprehensively analyze the fiber nonlinearity crosstalks between DAS and communication channels through numerical results and 40×800-Gb/s 90-km experimental demonstration. Our findings indicate that conventional pulse-based DAS is unsuitable for in-band DWDM coexistence system, whereas pulse-compression DAS shows negligible penalties with legacy coherent transceivers. \textcopyright2025 The Author(s)
        \vspace{-0.2cm}
    \end{ecoc_abstract}
\end{strip}


\section{Introduction}
\vspace{-0.1cm}
Future metro fiber networks will need real-time health monitoring, driving the development of integrated sensing and communication over fiber (ISACoF) \cite{huang2019first}. ITU-T recommends using 1625 nm or 1650 nm channels for in-service fiber monitoring \cite{ITUT}. Recently, ITU-T SG15 has also explored the coexistence of distributed fiber-optic sensing (DFOS) and communication, particularly using dense wavelength division multiplexing (DWDM) \cite{ITUTSG15}.

Several studies have explored the coexistence of distributed acoustic sensing (DAS) and communication data streams using WDM. Experiments inserting chirped-pulse DAS among 13 data wavelengths examined performance degradation in both co- and counter-propagation over 80 km fibers \cite{jia2021experimental}. Co-propagation of L-band DAS and 100-Gb/s coherent signals over 50 km showed no communication penalty with chirped-pulse DAS \cite{shen2022co}. The impact of stimulated Raman scattering between OTDR and other WDM channels suggests the need for pre-corrections in both OTDR and data channels \cite{zhang2022optical}. A field trial demonstrated real-time DAS and 400GbE coexistence, analyzing the tradeoff between DAS and data channel power \cite{huang2023field}. However, most prior work focuses on how much DAS degrades communication signals (e.g., BER or SNR), with little discussion on why and how DAS causes nonlinear crosstalk in legacy DWDM metro networks. Crucially, there is no systematic study on fiber nonlinearity crosstalk and tradeoffs needed before deploying DWDM ISAC networks.

This paper provides simulation and experimental analysis of fiber nonlinearity crosstalk between C-band 40$\lambda$ x 800ZR data channels and a DAS channel over 90 km, focusing on nonlinear phase and polarization noise (PPN) and the Raman effect. Three key findings are highlighted: (1) For nearby data channels (<0.8 THz), high-peak-power DAS pulses can induce strong nonlinear birefringence, causing rapid polarization rotation—often a bigger issue than cross-phase modulation (XPM), as depolarization occurs before phase noise recovery in legacy transceivers. (2) For data channels farther from the DAS wavelength (>0.8 THz), the main coexistence problem is the stimulated Raman effect (SRS), especially in ZR ($\ge$80 km) metro networks. The burst-like DAS pulses cause sudden power fluctuations in data channels, which can impact legacy amplifiers and receivers. (3) Pulse-compression-based DAS has minimal impact on data channels due to its low peak power and longer pulse duration.
\vspace{-0.1cm}
\section{Nonlinear effects in DWDM coexistence system}
\begin{figure}[t!]
    \centering
    \includegraphics[height=2.8cm]{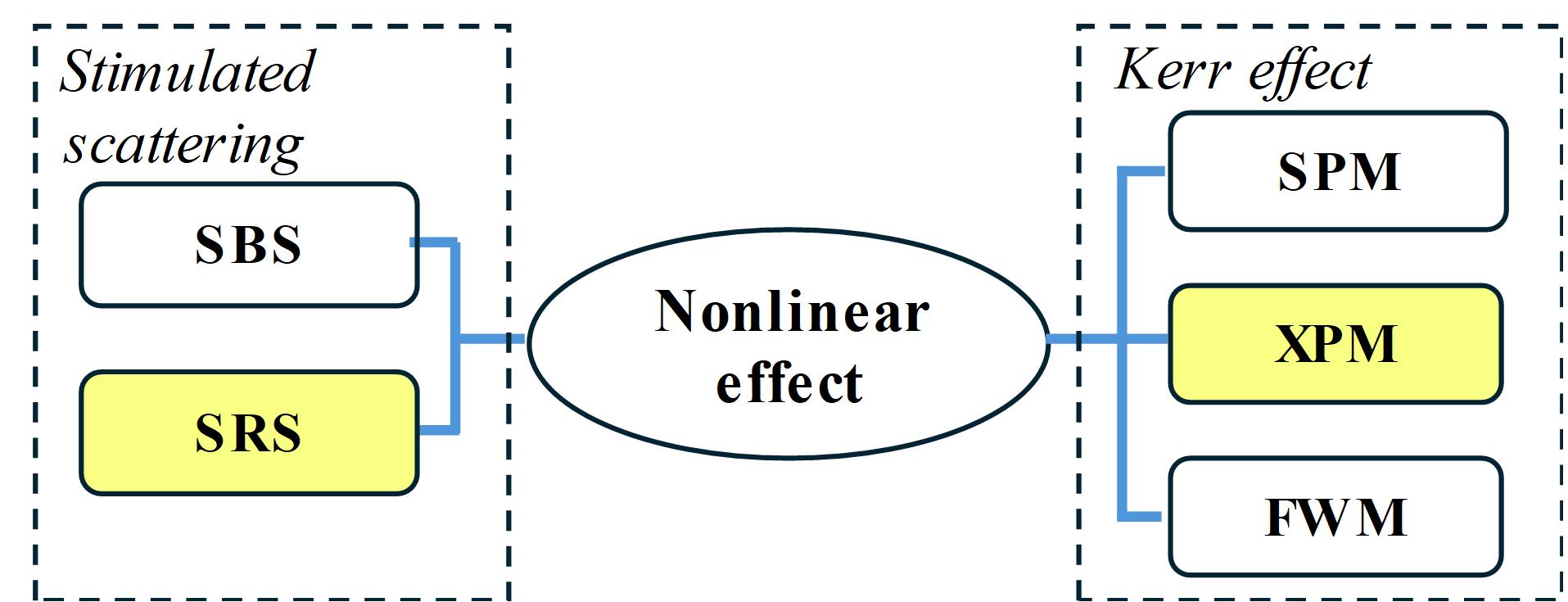}
    \caption{Nonlinear effects in optical fibers. }
    \label{fig:figure1}
    \vspace{-0.2cm}
\end{figure}
For metro link > 50 km, the nonlinear length is sufficient for averaging the rapid mixing of the polarization state on the Poincaré sphere \cite{wai1991stability}. We have
\begin{multline*}
\label{Eq1}
\frac{\partial u}{\partial z} = - \frac{\alpha}{2} u - \frac{i \beta_2}{2} \frac{\partial^2 u}{\partial t^2} \nonumber + i \cdot  \frac{8}{9} \gamma \Big [ (1 - f_R){\vert\vert u \vert\vert}^2 u \\
+ f_R \cdot u \int_{-\infty}^{\infty} h_R(\tau) \left( {\vert\vert u(t-\tau) \vert\vert}^2 \right) d\tau \Big ], \tag{1}
\end{multline*}
where $u(z, t)$ denotes $E_{x}$ or $E_{y}$ polarizations, $\beta_2$, $\gamma$, $f_R$ and $h_R$ are the group velocity dispersion and nonlinear parameter, Raman contribution factor and response function. Fiber nonlinearities are concluded in Fig. \ref{fig:figure1}, where highlighted blocks are the main nonlinear impacts induced by the DAS pulse. Because the minimum granularity of flex-grid DWDM is 12.5 GHz, the crosstalk by stimulated brilliouin scattering (SBS) can be ignored with the properties of narrow-band gain and typical 11.8 GHz brilliouin frequency shift (BFS).

\begin{figure}[t!]
    \centering
    \includegraphics[height=4cm]{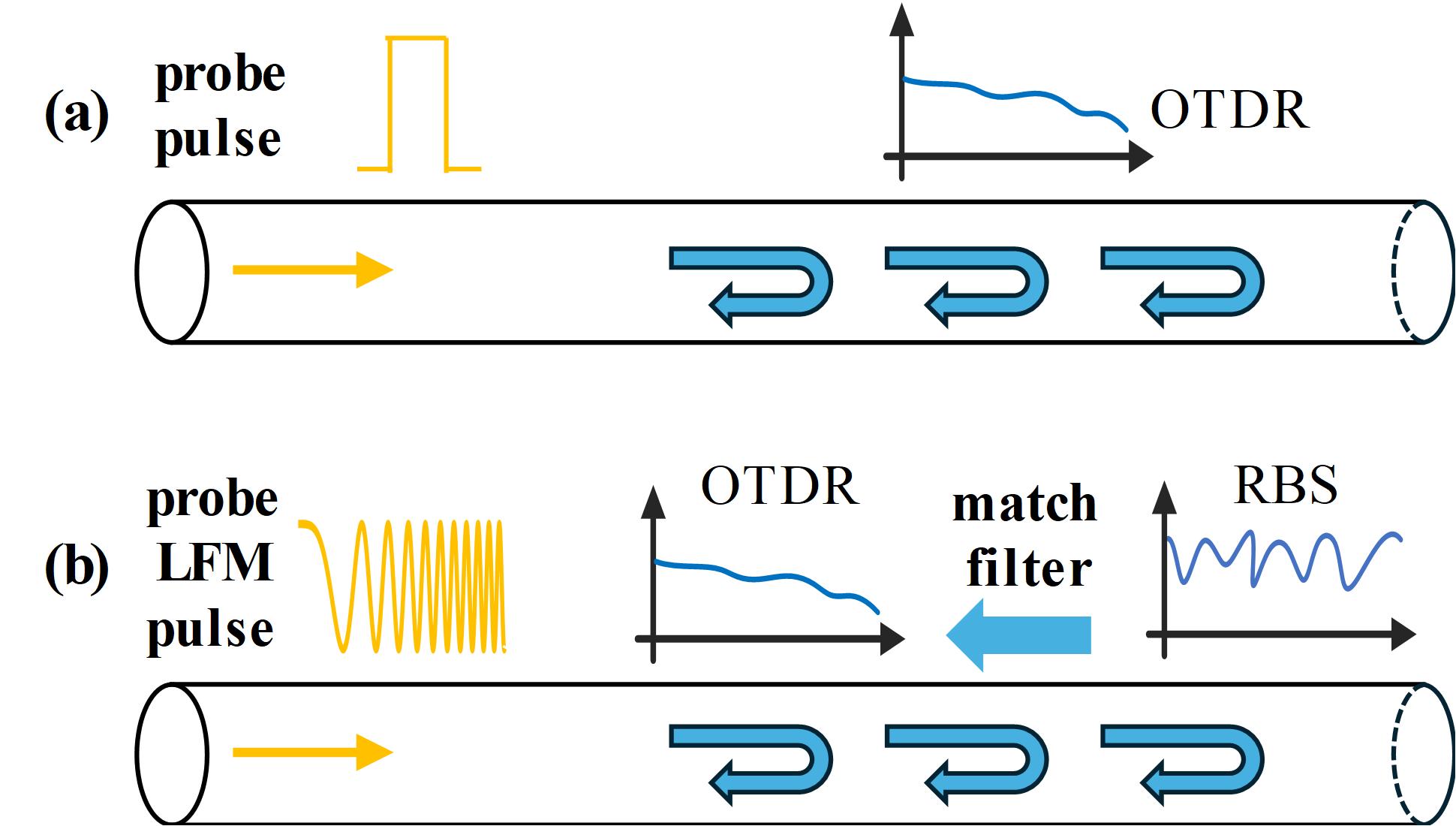}
    \caption{(a) Conventional pulse-based DAS and (b) LFM pulse-compression DAS}
    \vspace{-0.2cm}
    \label{fig:figure2}
\end{figure}

Fig. \ref{fig:figure2}(a) and (b) show conventional pulse-based DAS and linear frequency-modulated (LFM) pulse-compression DAS schemes. Conventional pulse-based DAS cannot cover ZR distances ($\geq$ 80 km) without heavy averaging or Raman amplification, so two bidirectional interrogators are needed for the 90-km fiber in our experiment. In contrast, LFM pulse-compression DAS, with longer pulses and matched filtering, can monitor up to 100 km \cite{r2019distributed}, meeting 800ZR requirements. Also, conventional DAS needs much higher peak power for long distances, greatly increasing nonlinear crosstalk to other channels.

\begin{table}[h!]
    \vspace{-0.2cm}
    \centering
    \caption{Numerical simulation parameters} \label{tab:Numerical}
    \begin{tabular}{|c|c|}
        \hline  \textbf{Parameter} & \textbf{Value} \\
        \hline  Fiber length & 90 km G.652D \\
        \hline  DAS channel & C34 \\
        \hline  Data channels & C21-C33 \& C35-C61 \\
        \hline  PMD coefficient & 0.1 ps/$\sqrt{km}$ \\
        \hline  Raman factor & 18 \% \\ 
        \hline  Raman response time & 3 fs \\
        \hline  Data per channel & 0 dBm 80 GBaud \\
        \hline  DAS pulse duration & 500 ns \\
        \hline  Pulse rise/fall time & 50 ns \\
        \hline  Peak power of DAS & 17-21 dBm \\
        \hline  SSFM stepsize & 5 m \\
        \hline
    \end{tabular}
    \vspace{-0.3cm}
\end{table}%

\begin{figure}[t!]
    \centering
    \includegraphics[height=3.5cm]{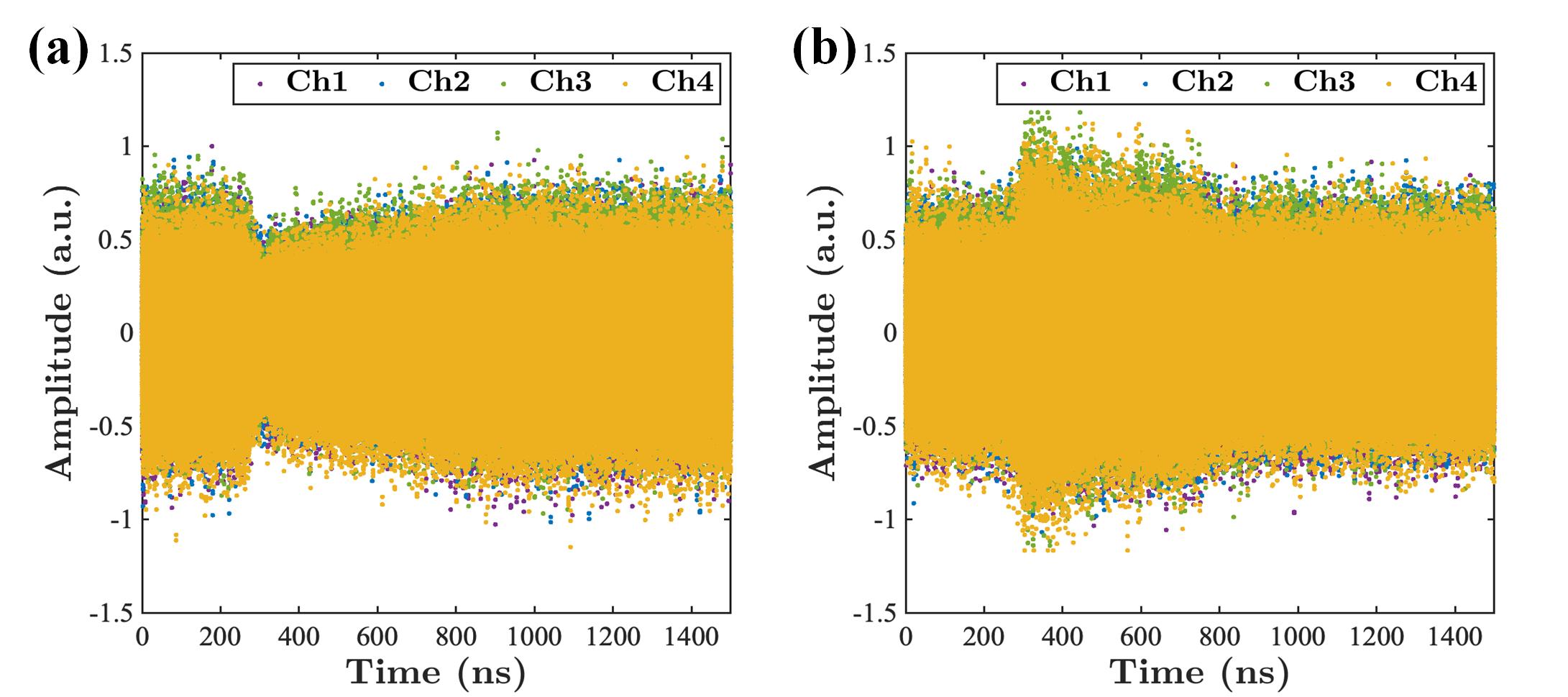}
    \caption{Pulse-induced waveform distortions of (a) C44 and (b) C24, burst power-fluctuating due to Raman effect. }
    \label{fig:figure3}
\end{figure}

\begin{figure}[t!]
    \centering
    \includegraphics[height=4.2cm]{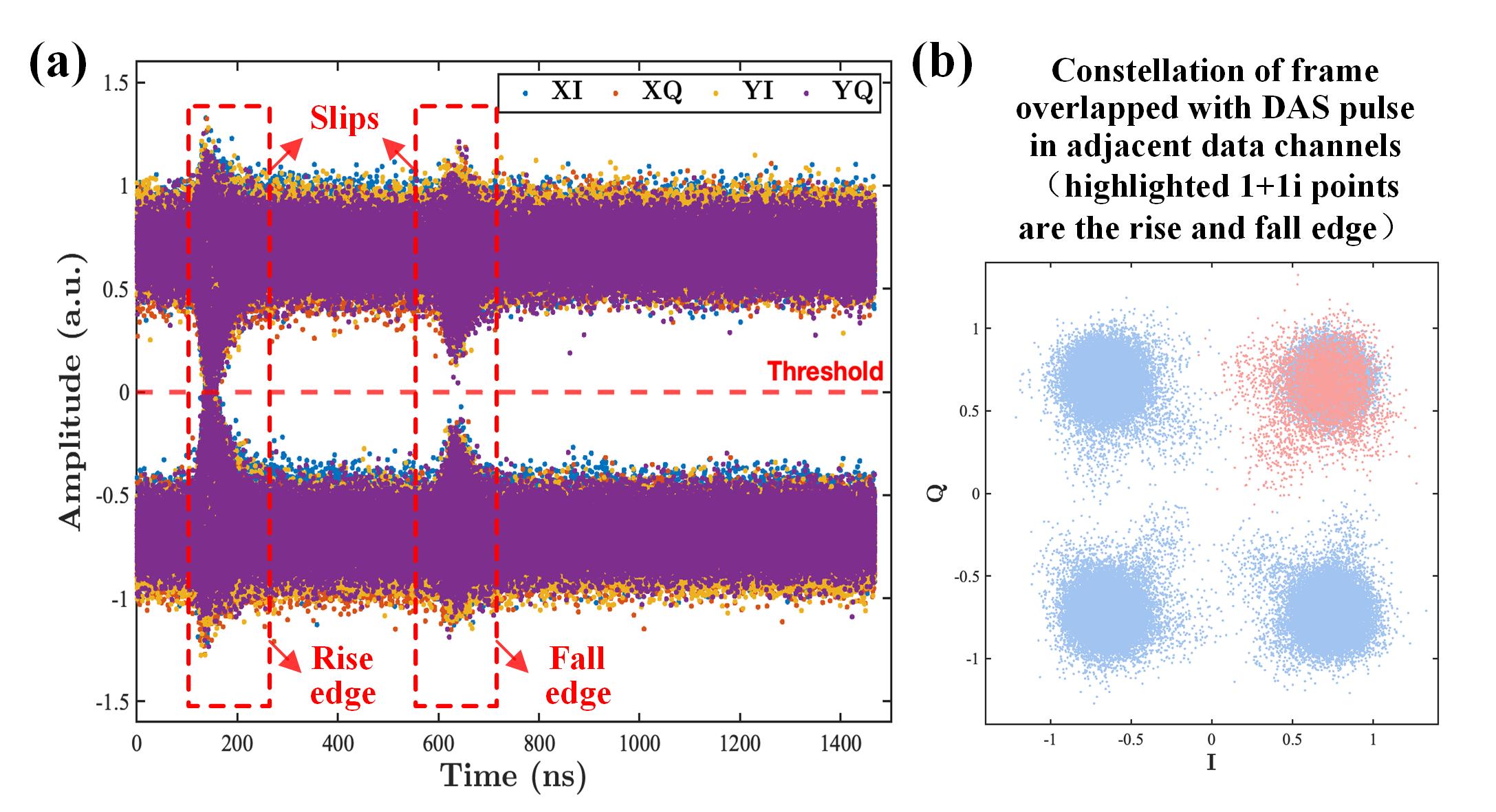}
    \caption{(a) Decoded data (C34) in 20 dBm peak power pulse (C33). (b) Constellation with rise/fall time highlighted.}
    \label{fig:figure4}
    \vspace{-0.1cm}
\end{figure}

We use the Manakov Equation (Eq. \ref{Eq1}) to simulate 90 km fiber coexistence between a pulse-based DAS channel at C34 and ITU-T C21–C61 communication channels (specifications in Table \ref{tab:Numerical}). All other channels transmit 80 GBaud QPSK data. To allow DAS interrogation from both ends over 50 km, we set a 500 ns pulse width and vary the peak power from 17 to 21 dBm—levels chosen to approach, but not cause, significant nonlinear effects \cite{martins2013modulation}. The DAS pulses also have 50 ns rise/fall egdes, matching commercial systems. For data channels more than 0.8 THz from the DAS channel in co-propagation, Fig. \ref{fig:figure3} shows power fluctuations in data frames overlapping with pulses, caused by the Raman effect. Fig. \ref{fig:figure4}a shows that adjacent channels (<0.8 THz) experience burst signal slips during the DAS pulse edges. In counter-propagation, data channels are unaffected. Fig. \ref{fig:figure4}b also shows diffused constellation points during rise/fall edges, confirming pulse-induced distortions from nonlinear polarization and XPM noise.
\vspace{-0.4cm}

\vspace{-0.2cm}
\section{Experimental setup and result discussions}
\begin{figure*}[t]
    \centering
    \includegraphics[height=5cm]{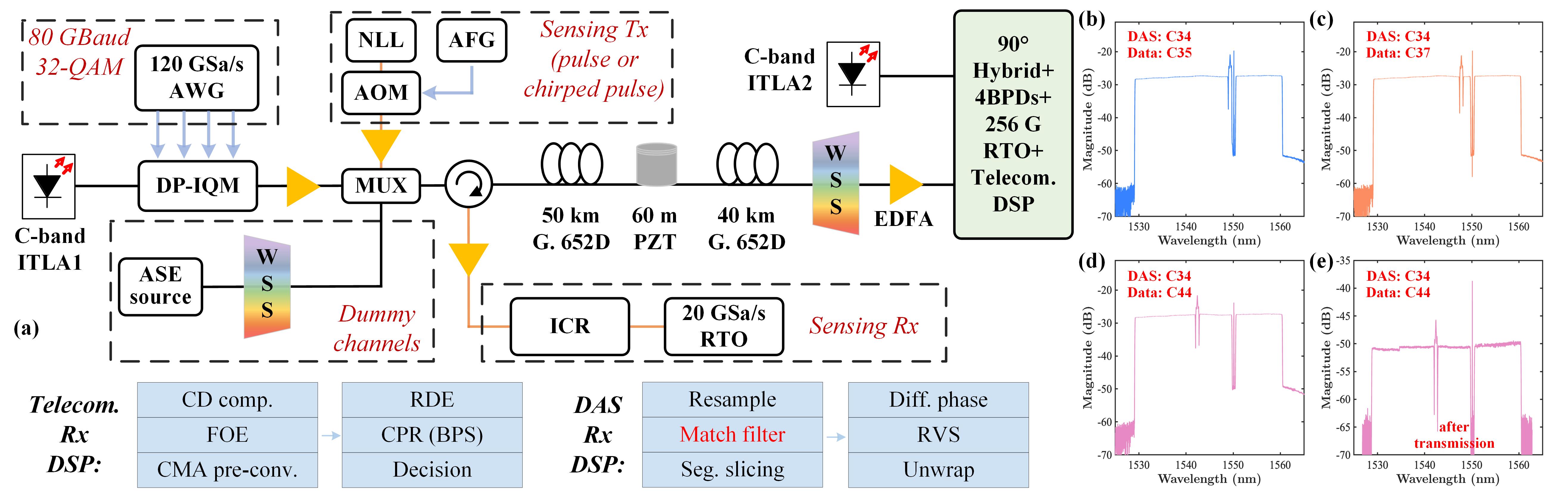}
    \caption{(a) Experimental setup of co-propagating case. (b)-(e) Spectrum of testing data channels, co-propagating with pulse. }
    \label{fig:figure5}
\end{figure*}
 Coexistence experiment of 90 km 40$\lambda$ x 80 GBaud 32/16/4-QAM channels and 1$\lambda$ DAS channel is consistent with numerical simulation. Both co-propagation and counter-propagation experiments are done, with only co-propagation casedepicted in Fig. \ref{fig:figure5}a.  We slide the wavelength of integrated tunable laser assembly (ITLA) to test the communication performance of each channel, while other data channels are loaded with amplified spontaneous emission (ASE) noise at same power level. Dummy band is generated by a C-band ASE source and shaped by a wavelength selective switch (WSS, Finisar 4000A). Communication setup consists of 120 GSa/s arbitary waveform generator (AWG, Keysight 8194) and 256 GSa/s real-time oscilloscope (RTO UXR0804). Sensing includes narrow linewidth laser (NLL, NKT X15) and acousto-optic modulator (AOM, G\&H) for generating pulse of 500 ns or LFM-pulse of 5000 ns and 15 MHz sweeping bandwidth. A 60 m piezoelectric transducer (PZT) acts as a vibrator between two fiber spools. An integrated coherent receiver (ICR, Lumentum Class 40) and 20 GSa/s scope (Keysight, MSOS404A) are used to receive the Rayleigh backscattering signal (RBS).
 \begin{table}[h!]
    \vspace{-0.1cm}
    \centering
    \caption{Raman effect induced burst power fluctuation. Define power variation exceeding 20\% of the average as fail.} 
    \label{tab:Raman}
    \begin{tabular}{|c|c|}
        \hline  \textbf{Pulse cases} & \textbf{Remark} \\
        \hline  peak power > 19 dBm, C21-C25   & fail \\
        \hline  peak power > 19 dBm, C43-C61 & fail \\
        \hline  C26-C33 \& C35-C42 & pass \\
        \hline  \textbf{LFM-pulse cases (C21-C61)} & \textbf{Remark} \\
        \hline \multicolumn{2}{|c|}{No fluctuations under peak power < 13 dBm} \\
        \hline 
    \end{tabular}
    \vspace{-0.2cm}
\end{table}%

 We analyze the Raman effect for co-propagation cases, with pulse and LFM-pulse results shown in Table \ref{tab:Raman}. Practically, burst power fluctuation exceeding the threshold will cause additional penalty to legacy amplifiers and ZR transceivers.
 
\begin{figure}[t!]
    \centering
    \includegraphics[height=4cm]{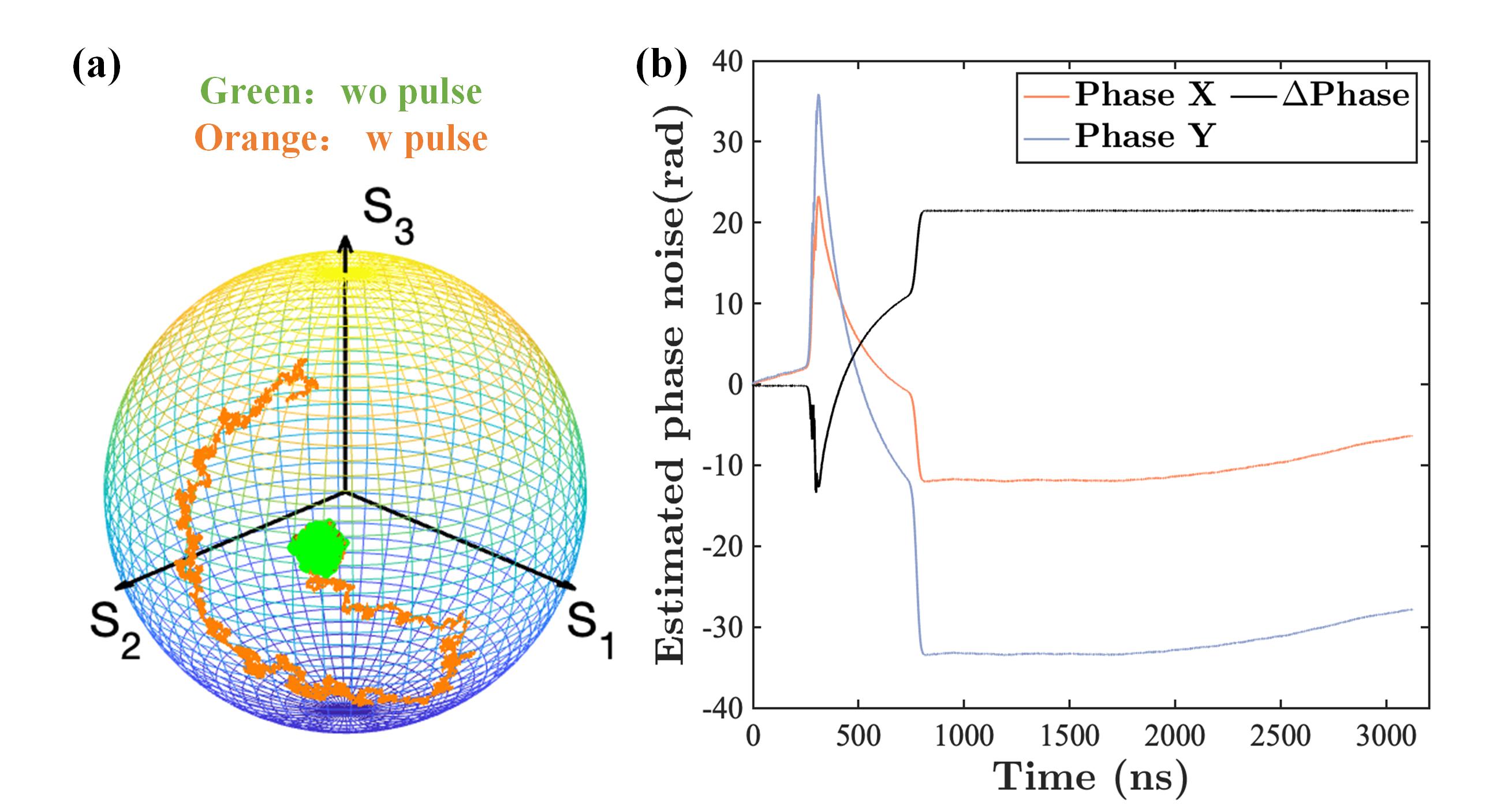}
    \caption{(a) Poincaré sphere from taps. (b) Phase noise plot.}
    \label{fig:figure6}
    \vspace{-0.1cm}
\end{figure}
Then we investigate the nonlinear polarization and phase noise of the adjacent channels (C26-C32 \& C35-C42). The Stokes parameters of polarization are obtained by equalization taps \cite{wang2025forward}. Fig. \ref{fig:figure6}a indicates a burst polarization rotation over the pulse duration when peak power exceeds 17 dBm. Fig. \ref{fig:figure6}b also reveals that different XPM-induced phase noise is observed of two polarizations with pulse on. These findings support that both nonlinear polarization and phase noise exist in the coexistence system due to the polarization misalignment of pulse and data. Furthermore, once we can successfully track the polarization change in the experiment, we can recover the XPM-induced phase noise, suggesting the former effect may be the main constraint of coexistence system.
\begin{figure}[t!]
    \centering
    \includegraphics[height=3.4cm]{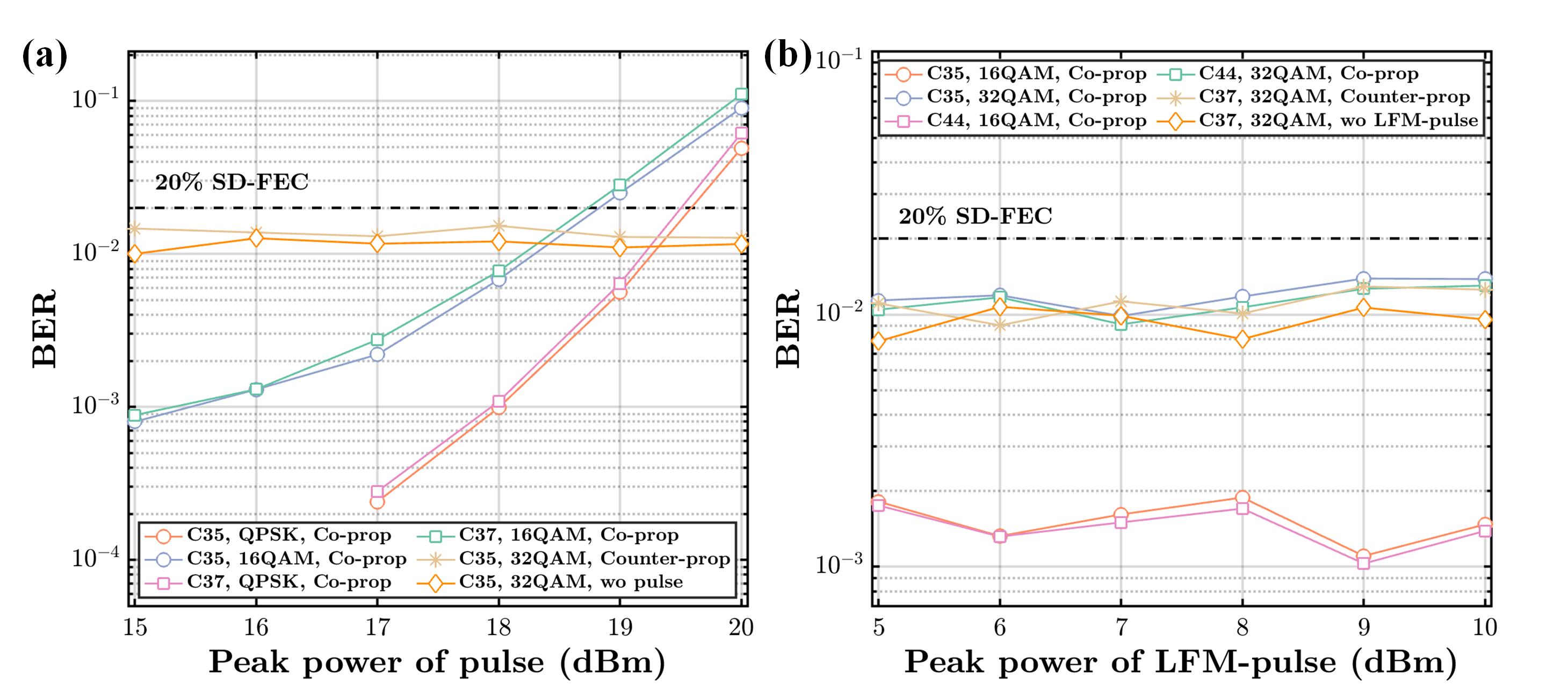}
    \caption{(a) BER with pulse. (b) BER with LFM-pulse.}
    \label{fig:figure7}
    \vspace{-0.3cm}
\end{figure}

To analyze the communication performance, we set the launch power of each data channel to be 0 dBm and scan the peak power of the co-propagation pulse from 15 dBm to 20 dBm. Fig. \ref{fig:figure7}a shows a sharp degradation of bit error rate (BER) as the peak power increases both in the 1st and 3rd adjacent channels. In contrast, negligible penalties are observed when data channels transmit accompanied with LFM-pulse from 5 dBm to 10 dBm, as shown in Fig. \ref{fig:figure7}b. Considering 16.7\% overhead of SD-FEC, we achieve $40*\rm{800G}*83\% = 26.7\, Tb/s$ ZR transmission.
\begin{figure}[t!]
    \centering
    \includegraphics[height=3.4cm]{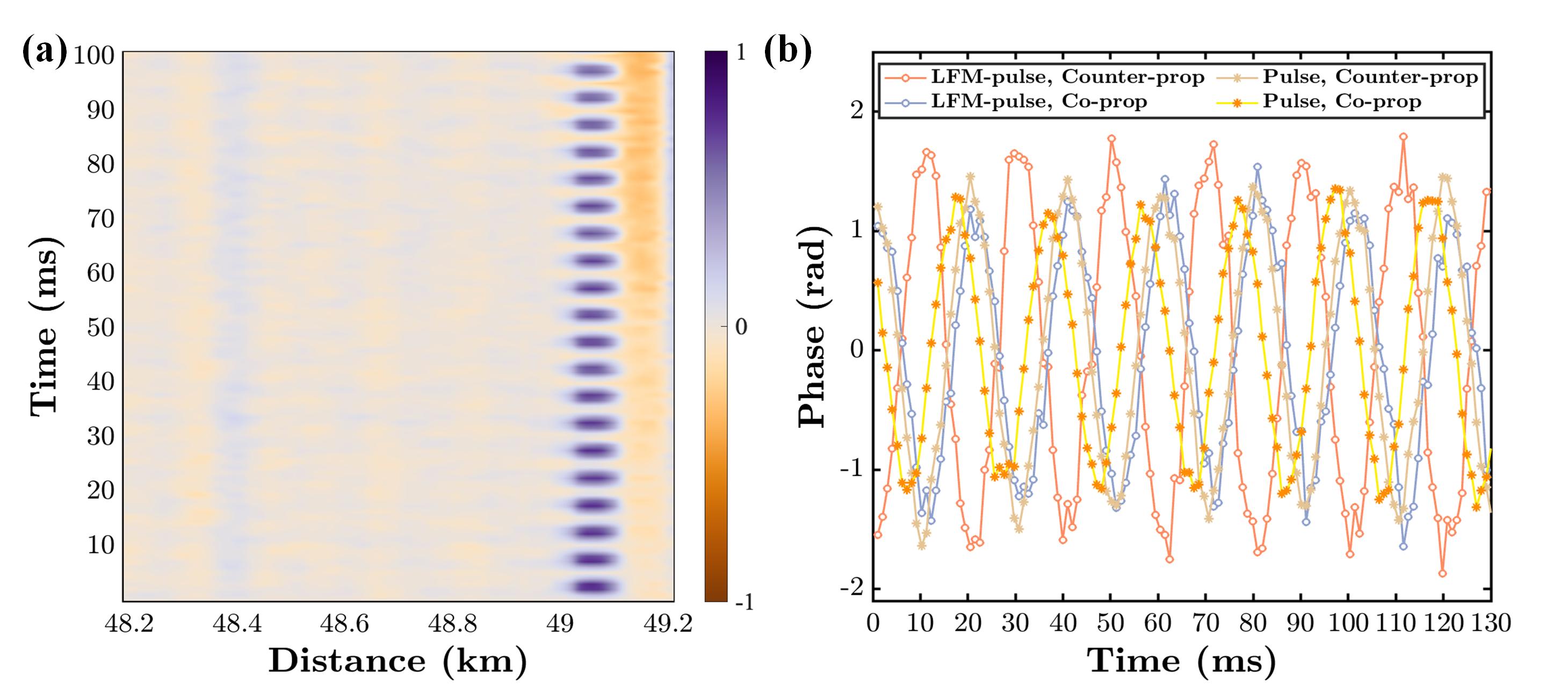}
    \caption{(a) Waterfall plot. (b) Demodulated vibrations.}
    \label{fig:figure8}
    \vspace{-0.2cm}
\end{figure}

Pulse (peak power 15-20 dBm) and LFM-pulse (peak power 5-10 dBm) reaches the similar SNR of DAS traces, while LFM-pulse can achieve finer sensing resolution. As pulse setup is hard to cover 90 km fiber without fadings, we need to do both co-propagation and counter-propagation from both sides. Fig. \ref{fig:figure8}a is the demodulated waterfall plot of 200 Hz vibration, interrogated by the co-propagation LFM pulse. Also, a 50 Hz vibration is successfully detected via co-/counter-propagation and pulse/LFM-pulse, shown in Fig. \ref{fig:figure8}b.
\vspace{-0.2cm}
\section{Conclusion}
We studied the in-band nonlinear crosstalks from conventional pulse-based and pulse-compression based DFOS in 800GZR metro networks. For data channels over 0.8 THz from the DAS channel (co-propagation), the main issue of coexistence is SRS-induced power fluctuation. For channels within 0.8 THz, nonlinear polarization rotation may have more severe impact on legacy coherent transceivers, compared with XPM-induced phase noise. Counter-propagating DAS causes little penalty to data channels, but one-side DAS cannot cover ZR distances. Besides, pulse-compression based DAS is highlighted for negligible nonlinear penalty to data channels and the ability to fully cover ZR distances. These findings on the nonlinear mechanisms and tradeoffs between sensing and communications will help guide future DFOS and high-speed communication coexistence design and standards.

\clearpage
\section{Acknowledgements}
This work is supported by the Hong Kong Research Grants Council (GRF 15227321, GRF 15225423) and Hong Kong Polytechnic University project 1-CD8L. We would like to acknowledge the contributions from Dr. Qirui Fan and Mr. Chuang Xu.

\defbibnote{myprenote}{%

}
\printbibliography[prenote=myprenote]

\vspace{-4mm}

\end{document}